\documentclass[conference]{IEEEtran}
\IEEEoverridecommandlockouts

\usepackage{silence} 
\usepackage{cite}
\usepackage{amsmath,amssymb,amsfonts}
\usepackage{algorithmic}
\usepackage{graphicx}
\usepackage{textcomp}
\usepackage{xcolor}
\usepackage{comment}
\usepackage{xurl}

\usepackage{booktabs}
\usepackage[acronym, automake]{glossaries}
\makeglossaries
\newacronym{PV}{PV}{photovoltaic}
\newacronym{DER}{DER}{distributed energy resource}
\newacronym{DLNN}{DLNN}{deep learning neural network}
\newacronym{AMI}{AMI}{advanced metering infrastructure}
\newacronym{NN}{NN}{neural network}
\newacronym{BNN}{BNN}{Bayesian neural network}
\newacronym{DSSE}{DSSE}{distribution system state estimation}
\newacronym{WLS}{WLS}{weighted least squares}
\newacronym{DN}{DN}{distribution network}
\newacronym{DSO}{DSO}{distribution system operator}
\newacronym{LV}{LV}{low-voltage}
\newacronym{MV}{MV}{medium voltage}
\newacronym{EV}{EV}{electric vehicle}
\newacronym{SM}{SM}{smart meter}
\newacronym{QR}{QR}{quantile regression}
\newacronym{RMSE}{RMSE}{root mean square error}
\newacronym{SE}{SE}{state estimation}
\newacronym{PI}{PI}{prediction interval}
\newacronym{LVSE}{LVSE}{low-voltage state estimation}
\newacronym{FS}{FS}{feature set}
\newacronym{KL}{KL}{Kullback-Leibler}
\newacronym{ICT}{ICT}{information and communication technology}
\newacronym{FM}{FM}{flexibility market}
\newacronym{BRP}{BRP}{balance responsible party}
\newacronym{SCADA}{SCADA}{supervisory control and data acquisition}
\newacronym{mPMU}{$\mu$PMU}{micro-phasor measurement unit}
\newacronym{PLC}{PLC}{programmable logic controller}
\newacronym{RTU}{RTU}{remote terminal unit}
\newacronym{HEMS}{HEMS}{home energy management system}
\newacronym{IED}{IED}{intelligent electronic device}
\newacronym{HP}{HP}{heat pumps}
\newacronym{CPS}{CPS}{cyber-physical system}

\def\BibTeX{{\rm B\kern-.05em{\sc i\kern-.025em b}\kern-.08em
    T\kern-.1667em\lower.7ex\hbox{E}\kern-.125emX}}
\begin{document}

\title{Threat Scenarios and Monitoring Requirements for Cyber-Physical Systems of Flexibility Markets
\thanks{This work is funded by the Innovation Fund Denmark under File No. 91363 and the Swedish Energy Agency as part of the ERA-Net project HONOR.}
}

\author{

\IEEEauthorblockN{Nils Müller, Kai Heussen}
\IEEEauthorblockA{\textit{Wind and Energy Systems Department} \\
\textit{Technical University of Denmark}\\
Lyngby, Denmark \\
\{nilmu; kheu\}@dtu.dk}

\and

\IEEEauthorblockN{Zeeshan Afzal, Mathias Ekstedt}
\IEEEauthorblockA{\textit{Department of Computer Science} \\
\textit{KTH Royal Institute of Technology}\\
Stockholm, Sweden \\
\{zafzal; mekstedt\}@kth.se}

\and

\IEEEauthorblockN{Per Eliasson}
\IEEEauthorblockA{\textit{Foreseeti AB} \\
Stockholm, Sweden \\
per.eliasson@foreseeti.com}

}

\IEEEoverridecommandlockouts


\maketitle

\begin{abstract}
The ongoing integration of renewable generation and distributed energy resources introduces new challenges to distribution network operation. Due to the increasing volatility and uncertainty, \glspl{DSO} are seeking concepts to enable more active management and control. \Glspl{FM} offer a platform for economically efficient trading of electricity flexibility between \glspl{DSO} and other participants.
The integration of cyber, physical and market domains of multiple participants makes \glspl{FM} a system of \glspl{CPS}. While cross-domain integration sets the foundation for efficient deployment of flexibility, it introduces new physical and cyber vulnerabilities to participants. 
This work systematically formulates threat scenarios for the \glspl{CPS} of \glspl{FM}, revealing several remaining security challenges across all domains.
Based on the threat scenarios, unresolved monitoring requirements for secure participation of \glspl{DSO} in \glspl{FM} are identified, providing the basis for future works that address these gaps with new technical concepts.

\end{abstract}

\begin{IEEEkeywords}
distribution grids, flexibility markets, threat scenarios, monitoring requirements, cyber-physical power systems
\end{IEEEkeywords}

\glsresetall
\section{Introduction} \label{sec:introduction}
To reach the European goal of carbon neutrality in 2050, electricity generation and consumption must undergo radical changes. 
While the share of renewable generation needs to increase, electrification through devices such as \glspl{EV} and \glspl{HP} will drive up and reshape electricity demand. 
This extensive installation of \glspl{DER} will introduce more uncertainty and volatility, which radically changes usage of \glspl{DN}, potentially requiring expensive grid reinforcements. 
A widely discussed alternative is the use of end user flexibility, referred to as demand response \cite{spiliotis2016demand}. 
By reducing equipment loading at peak hours, \glspl{DSO} can use local flexibility to delay or avoid investments for reinforcement of transformers and power lines. 

As a framework for the integration of local flexibility, a widely promoted approach are \glspl{FM} \cite{jin2020local}. \glspl{FM} constitute a competitive trading platform for electricity flexibility in a geographically restricted area such as towns \cite{olivella2018optimization}. 
A typical setup of market participants consists of a \gls{DSO}, a \gls{BRP}, several aggregators and a market operator. 
Aggregators pool and manage multiple small residential flexibility assets.
In this way, they enable end users to participate in \glspl{FM}.
\glspl{DSO} and \glspl{BRP} typically are flexibility buyers, while aggregators constitute sellers. 
\glspl{DSO} procure flexibility for operational purposes, such as congestion management or voltage control. 
\glspl{BRP} buy flexibility for portfolio optimization. 
By adjusting power demand of aggregated flexibility assets, aggregators make profits according to flexibility contracts. 
Owners of flexibility assets earn profits by providing \glspl{DER}, such as \glspl{HP} or \glspl{EV}, to aggregators. 

The foundation of a \gls{FM} is a strong integration of cyber, physical and market domains of multiple actors, making it a system of \glspl{CPS}. 
While this cross-domain integration sets the foundation for efficient deployment of flexibility assets, it introduces new vulnerabilities to involved participants and their systems. 
By applying end user flexibility to avoid critical grid states, \gls{DSO} grid operation becomes partly dependent on third parties. 
Moreover, the required use of \gls{ICT}, including less secure public networks, and the strong coupling with the physical and market domain opens doors for cyber criminals, aiming at social or financial damage. 
In addition, incorporating home devices of end users as flexibility assets also requires transmission, storage and processing of sensitive data. 

This paper contributes to the identification and analysis of possible risks and security requirements in the \glspl{CPS} of \glspl{FM}.
The work first provides an overview of possible threat scenarios, which result from a comprehensive and original system analysis. 
Thereafter, unresolved monitoring requirements for secure participation of \glspl{DSO} in \glspl{FM} are derived from the threat scenarios.
Objective of this work is to provide a foundation and motivation for future works addressing the identified gaps with new technical concepts and case studies.

\subsection{Related work}
As highlighted by \cite{alizadeh2016flexibility} and \cite{sperstad2020impact}, the influence of flexibility on power system security constitutes a research gap, as most existing works focus solely on benefits of flexibility usage.
Some works shed light on specific physical threats, such as uncertain customer behavior \cite{zeng2017impact} or financial threats, e.g. financial risk due to the intermittent nature of flexibility assets \cite{ghose2019risk}.
Other works such as \cite{chehri2021security} and \cite{smart-meter} investigate cyber threats introduced by the application of new smart grid technologies, including \glspl{SM} and advanced metering infrastructure. 
In \cite{enisa}, a number of cyber threats are identified and mapped to grid assets and threat agents. 
The work also addresses possible security controls to reduce exposure to threats.
However, these works only focus on particular threats or threat categories and do not specifically address \glspl{FM}.

In \cite{sperstad2020impact} possible positive and negative impacts of flexibility on the security of supply are discussed from a physical and a cyber perspective. 
A major physical threat is seen in the rebound effect of flexibility activations which may shift load peaks, and results in even more severe situations. 
A flexibility-induced cyber threat is seen in load-altering attacks that may impact the bulk power system without compromising better protected assets on transmission level. 
To the best of the authors knowledge, \cite{sperstad2020impact} is the only work that provides cyber and physical threat scenarios in the context of flexibility. 
However, as threat scenarios are no major concern of the work, it does not provide a comprehensive and systematic overview. 
Moreover, it neither takes characteristics of \glspl{FM} into account nor derives unresolved security requirements to motivate new research directions for future studies. 

\subsection{Contribution and paper structure}
The contribution of this paper is twofold:
\begin{itemize}
    \item Systematic formulation of threat scenarios for the \glspl{CPS} of \glspl{FM}. Scenarios result from an original system analysis and consider origins in various domains, emphasizing the interaction among the cyber, physical and market domain.
    \item Identification of unresolved monitoring requirements for \glspl{DSO} participating in \glspl{FM} as foundation for new technical concepts and case studies addressing these gaps.
\end{itemize}

The remainder of this paper is structured as follows. 
Section \ref{sec:threat_scenarios} provides a systematic overview of threat scenarios for the \glspl{CPS} of \glspl{FM}. 
Section \ref{sec:security_requirements} identifies monitoring requirements for participation of \glspl{DSO} in \glspl{FM}. Finally, Section \ref{sec:conclusion} concludes the paper. 
\section{Threat scenarios} \label{sec:threat_scenarios}
This section is concerned with the systematic formulation of threat scenarios for the \glspl{CPS} of \glspl{FM}. Subsection \ref{subsec:threat_scenario_formulation} presents the scenario formulation approach, followed by scenario descriptions in Subsections \ref{subsec:SM} to \ref{subsec:vendor}.

\subsection{Threat scenario formulation} \label{subsec:threat_scenario_formulation}
\begin{figure}[t]
\centering
\includegraphics[width=0.99\linewidth,trim=18 24.5 22 24.5,clip]{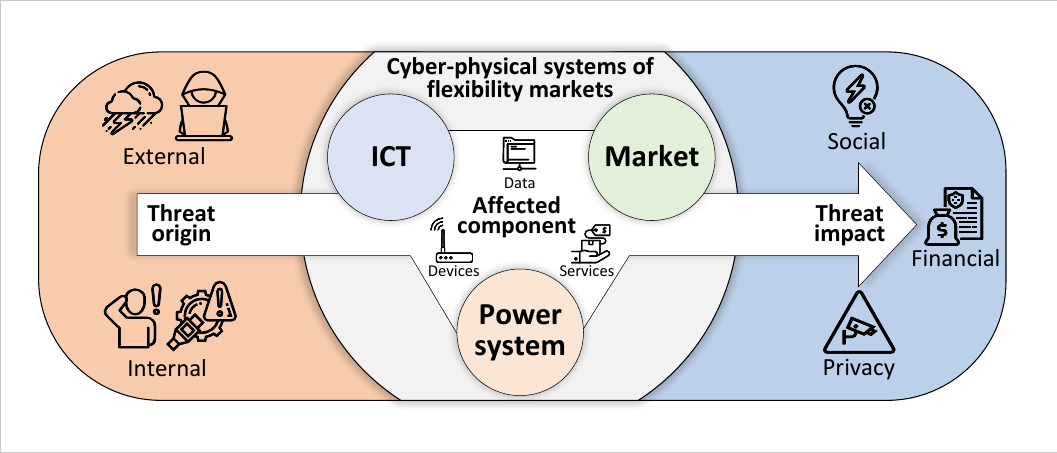}
\caption{Schematic representation of the threat scenario formulation.}
\label{fig:threat_scenario_formulation}
\end{figure}

To describe and compare scenarios with various backgrounds, a domain-neutral formulation is required, which still captures key information. Fig. \ref{fig:threat_scenario_formulation} represents the applied formulation concept. Threat origin, affected component and threat impact are selected as domain-independent key information.
The threat origin comprises two groups, namely external and internal. 
In Table \ref{tab:threat_origin} the considered origins are listed and allocated to one of the two groups, supplemented by information on their background. 
Table \ref{tab:threat_origin} indicates the broad spectrum of origins, enabling a holistic threat scenario investigation.

Typically, a critical situation develops around a specific component or component type in a system. 
A cyber attacker will most likely try to manipulate a specific data stream or device to launch the attack.
A famous example is the Industroyer malware attack on the Ukrainian power system, which targeted the control of circuit breakers in substations \cite{8220480}. 
An aggregator who falsely determines the potential of its portfolio will affect a flexibility service traded in the market. 

At the end of every scenario there is a potential negative impact, typically of social or financial nature. 
However, other impact, such as a loss of private information, are taken into account.
To allow for a better overview and to demonstrate how fundamentally different threat origins can result in similar critical situations, scenarios are grouped by the affected component. 
Table \ref{tab:summary_of_threat_scenarios} summarizes the scenarios, including threat origins, given as numbers referring to Table \ref{tab:threat_origin}, and impacts.
\setlength{\tabcolsep}{7pt}
\begin{table}[b]
\centering
  \caption{Overview and description of threat origins.}
  \label{tab:threat_origin}
  \begin{tabular}{llll}
    \hline
    \textbf{Nr.} & \textbf{Threat origin} & \textbf{Group} & \textbf{Background}\\
    \hline
    1& Device failure & Internal  & accidentally \\
    2& Human error & Internal & accidentally \\
    3& Market actors & Internal  & financial gain \\ 
    4& Insiders  & Internal & dissatisfaction \\     
    5& Consumer behavior & Internal & randomness \\
    6& Weather & External  & \begin{tabular}[c]{@{}l@{}} volatility, randomness\end{tabular} \\
    7& Price signals & External  & \begin{tabular}[c]{@{}l@{}} volatility, randomness \end{tabular}\\
    8& \begin{tabular}[c]{@{}l@{}}Organized cyber-criminals\end{tabular} & External & financial gain \\
    9& \begin{tabular}[c]{@{}l@{}}State-sponsored actors\end{tabular}& External & political \\
  \hline
\end{tabular}
\end{table}

\subsection{SM-based scenarios} \label{subsec:SM}
\Glspl{SM} may provide data to aggregators for flexibility planning and verification. Meters typically use a \gls{PLC} interface to communicate with the utility and have capability to remotely switch power on or off.

\subsubsection{Unauthorized access and modification of \gls{SM} data} \label{subsubsec:SM_modification}
Cyber-criminals could gain access to sensitive meter data such as consumption, credentials, and firmware information by exploiting known software vulnerabilities or by decrypting \gls{PLC} communication that uses weak encryption.
Additionally, if cyber-criminals take over the meter communication with the aggregator (e.g., by using the encryption key), they can send wrong consumption data. The asset owner may be fined for breaking contracts and removed from the portfolio. 
Having hold of private data and options for financial damage, cyber-criminals may aim at blackmailing flexibility asset owners. 

\subsubsection{Accessing and controlling multiple SMs} \label{subsubsec:multiple_SM_modification}
As extension of Scenario \ref{subsubsec:SM_modification}, state-sponsored actors may aim at accessing and controlling multiple \glspl{SM}. 
Attackers may use weaknesses of \glspl{SM}, such as static encryption keys. Typically, \glspl{SM} deployed by a \gls{DSO} share the same encryption key.
Thus, if attackers gain access to the encryption key of one \gls{SM}, they are able to extract useful information, such as the energy consumption behavior, of entire neighborhoods. 
At this stage, attackers can also take over remote on/off switching.
In \cite{alanazi2021load}, it is demonstrated how attackers can cause line trips through load oscillations by exploiting the switching capability of multiple \glspl{SM}. 
The result may be power outages, resulting in social and financial cost. 
Another attack path to target multiple \glspl{SM} is to launch a remote or physical attack on meter data concentrators. 



\subsection{Local controller-based scenarios}
Various actors of \glspl{FM} rely on local controllers. By interfacing the cyber and physical domain, they constitute critical system components, introducing potential threats.

\subsubsection{Modification of substation controller} \label{subsubsec:substation_controller_modification}
Primary and some secondary substations are equipped with controllers, such as \glspl{PLC}, \glspl{RTU} and \glspl{IED}. 
State-sponsored actors could place infected rootkits onto one or multiple local controllers.
By sending malicious control signals to circuit breakers and protection relays attackers could damage grid facility and disconnect customers. 
To hide the attack, normal operation values could be returned to the central control room. 
In \cite{liu2011false}, it is demonstrated that attackers can create such false data that will not raise an alarm by existing algorithms for bad data detection. 

\subsubsection{Modification of flexibility asset controller} \label{subsubsec:flexibility_asset_controller_modification}
Flexibility assets such as \glspl{EV} or \glspl{HP} are often controlled by \glspl{HEMS}.
These internet-connected systems typically have remote control capability and are based on off-the-shelf soft- and hardware, making them vulnerable to cyber attacks and asset owner modification. 
Flexibility asset owners may aim at financial gain by modifying setpoint boundaries or increasing setpoints before a service is activated, which manipulates the flexibility service of the aggregator.
Organized cyber-criminals could make use of weak password security and encryption to gain access to individual \glspl{HEMS}.
To blackmail customers, attackers may collect sensitive data, change setpoints to impair customer comfort and degrade flexibility assets, or increase costs by raising consumption or mitigating contracted flexibility activations. 
State-sponsored actors could infiltrate local controllers of multiple small or individual large flexibility assets. 
By changing the setpoints or switching assets on or off, they could introduce load peaks or oscillations to trigger transformer protection, resulting in customer disconnection. 
A coordinated attack on flexibility assets and grid protection mechanisms may result in severe physical damage of grid facilities and blackouts.   

\subsubsection{Failure of large flexibility asset controller} \label{subsubsec:flex_asset_controller_failure}
The activation of large flexibility assets may fails due to soft- or hardware failures. Compared to defects of small assets, the impact may be severe. An industrial plant could provide flexibility by reducing production capacity during times of high \glspl{EV} charging. Under these conditions, an activation failure could lead to congestion at the transformer. 
\setlength{\tabcolsep}{0.7pt}
\begin{table}[ht]
\centering
  \caption{Summary of the identified threat scenarios.}
  \label{tab:summary_of_threat_scenarios}
  \begin{tabular}{llll}
    \hline
    \textbf{Nr.} & \textbf{Threat scenario} &\textbf{Impact} & \textbf{Origin}\\
    \hline
    
    \multicolumn{4}{c}{\textbf{SM-based scenarios}}\\
\hline
1 & \begin{tabular}[c]{@{}l@{}}Unauthorized access \& modification of \gls{SM} data \end{tabular} & \begin{tabular}[c]{@{}l@{}} privacy \end{tabular} & 3,8\\
2 & \begin{tabular}[c]{@{}l@{}}Accessing and controlling multiple \glspl{SM} \end{tabular} & \begin{tabular}[c]{@{}l@{}} social,\,privacy \end{tabular} & 8,9\\
\hline

    \multicolumn{4}{c}{\textbf{Local controller-based scenarios}}\\
\hline
3 & \begin{tabular}[c]{@{}l@{}} Modification of substation controller \end{tabular} & \begin{tabular}[c]{@{}l@{}} social,\,financial \end{tabular} &  9 \\
4 & \begin{tabular}[c]{@{}l@{}} Modification of flexibility asset  controller \end{tabular} & \begin{tabular}[c]{@{}l@{}} social,\,privacy \end{tabular} &  3,8,9\\
5 & \begin{tabular}[c]{@{}l@{}} Failure of large flexibility asset controller \end{tabular} & \begin{tabular}[c]{@{}l@{}} social,\,financial \end{tabular} &  1\\
\hline

    \multicolumn{4}{c}{\textbf{Flexibility activation signal-based scenarios}}\\
\hline
6 & \begin{tabular}[c]{@{}l@{}} Tamper or disrupt flexibility activation signals \end{tabular} & \begin{tabular}[c]{@{}l@{}} social,\,financial \end{tabular} &  4,8,9\\
7 & \begin{tabular}[c]{@{}l@{}} Unintentional wrong activation of flexible assets \end{tabular} & \begin{tabular}[c]{@{}l@{}} social,\,financial \end{tabular} &  2\\
8 & \begin{tabular}[c]{@{}l@{}} Parallel flexibility activations with opposing or  \\   reinforcing effects \end{tabular} & \begin{tabular}[c]{@{}l@{}} financial \end{tabular} &  3,7\\
\hline

    \multicolumn{4}{c}{\textbf{Historical data-based scenarios}}\\
\hline
9 & \begin{tabular}[c]{@{}l@{}} Compromised data on \gls{DSO} or aggregator data \\  historian \end{tabular} & \begin{tabular}[c]{@{}l@{}} financial \end{tabular} &  1,2,4,9\\
\hline

    \multicolumn{4}{c}{\textbf{Flexibility request-based scenarios}}\\
\hline
10 & \begin{tabular}[c]{@{}l@{}} High uncertainty in the determination of flexi-\\bility needs \end{tabular} & \begin{tabular}[c]{@{}l@{}} financial \end{tabular} & 5,6,7 \\
11 & \begin{tabular}[c]{@{}l@{}} Uncertainty about power system states due to\\  frequent flexibility activations \end{tabular}& \begin{tabular}[c]{@{}l@{}} social,\,financial \end{tabular} & 3 \\
12 & \begin{tabular}[c]{@{}l@{}} Parallel events resulting in sudden change of  \\  flexibility needs \end{tabular} & \begin{tabular}[c]{@{}l@{}} social,\,financial \end{tabular} &  1,2,7,9 \\
\hline

    \multicolumn{4}{c}{\textbf{Flexibility offer-based scenarios}}\\
\hline
13 & \begin{tabular}[c]{@{}l@{}} Place wrong flexibility offers on the \gls{FM} \end{tabular} & \begin{tabular}[c]{@{}l@{}} social,\,financial \end{tabular} &  4,9\\
14 & \begin{tabular}[c]{@{}l@{}} High uncertainty in the determination of flexi-\\bility offers \end{tabular} & \begin{tabular}[c]{@{}l@{}} financial \end{tabular} &  5,6\\
\hline

    \multicolumn{4}{c}{\textbf{Flexibility measurement or schedule-based scenarios}}\\
\hline
15 & \begin{tabular}[c]{@{}l@{}} Disrupt or manipulate flexibility measurements  \\ and schedules \end{tabular} & \begin{tabular}[c]{@{}l@{}} financial,\,privacy \end{tabular} &  3,8\\
\hline

    \multicolumn{4}{c}{\textbf{Flexibility asset-based scenarios}}\\
\hline
16 & \begin{tabular}[c]{@{}l@{}} Unavailability of flexibility assets \end{tabular} & \begin{tabular}[c]{@{}l@{}} financial \end{tabular} &  1,5,8,9\\
\hline

    \multicolumn{4}{c}{\textbf{Vendor soft- and hardware-based scenarios}}\\
\hline
17 & \begin{tabular}[c]{@{}l@{}} Compromise vendor software and systems \end{tabular} & \begin{tabular}[c]{@{}l@{}} social,\,financial \end{tabular} &  9 \\

    \hline
  \end{tabular}
\end{table}
\subsection{Flexibility activation signal-based scenarios}
Flexibility activation signals comprise activation requests from flexibility buyers to sellers, and activation signals from aggregators to small flexibility assets. 
The transmission is typically conducted via public networks.

\subsubsection{Tamper or disrupt flexibility activation signals} \label{subsubsec:modification_of_activation_signals}
Aggregator employees may launch insider attacks, such as sending activation signals at wrong times or preventing required flexibility activations. 
Insiders of \glspl{DSO} may send wrong activation requests.
Flexibility activation signals could also be manipulated by cyber-criminals or state-sponsored actors through false data injection attacks, exploiting insecure authentication or weak encryption. 
Attackers could also flood flexibility assets with activation signals to disrupt the activation process.
In all these scenarios, attackers could prevent or temper required flexibility activations to leave congestions or voltage violations unresolved or even intensify them. 
Moreover, attackers could initiate critical grid states by activating flexibility assets. 
In both cases high social and financial costs are likely.  

\subsubsection{Unintentional wrong activation of flexibility assets} \label{subsubsec:unintentional_wrong_activation}
Human errors of various actors, such as \glspl{DSO} or aggregators, and in different process steps, from determining flexibility needs or potential to preparing and sending activation requests, could initiate wrong flexibility activations. 
Equivalent to intentional attacks, damage could be of social and financial nature.

\subsubsection{Parallel flexibility activations with opposing or reinforcing effects} \label{subsubsec:parallel_activations}
Flexibility services can be requested by different actors with distinct purposes. While a \gls{DSO} may intend to prevent congestion, a \gls{BRP} aims at portfolio optimization. 
Thus, flexibility services with opposing effects could be activated simultaneously, resulting in financial damage as services may be procured without achieving the desired outcome.
At the same time, price-based demand response introduces additional flexibility activations in \glspl{DN}. 
\Glspl{DSO} might be unaware of future behavior of price-driven loads during flexibility planning.
Thus, a risk for network violations exists if the \gls{DSO} service is reinforced by price-driven flexibility.

\subsection{Historical data-based scenarios}
Historical data is of high importance for several actors in \glspl{FM}. Threats emerge from potential data loss or manipulation. 

\subsubsection{Compromised data on DSO or aggregator data historian} \label{subsubsec:modified_historical_data}
Historical data provide necessary information for flexibility planning, activation and verification.
Typically, they are not checked for integrity, after being stored. 
However, integrity could be affected by human and transfer errors or attacks. 
Model development based on compromised data will weaken performance or might render models useless.
Financial damage may result due to imprecise flexibility planning and verification. 
In severe cases, power system monitoring techniques may fail, leaving critical grid conditions unresolved.

\subsection{Flexibility request-based scenarios}
To procure flexibility, \glspl{DSO} and \glspl{BRP} submit flexibility requests to the \gls{FM}. Depending on the market concept, requests can be formulated from intraday to months ahead.

\subsubsection{High uncertainty in the determination of flexibility needs} \label{subsubsec:Flex_demand_uncertainty}
\Glspl{DN} face increasing volatility due to the dependency of distributed energy resources on weather, consumer behavior and price signals. 
At the same time, \gls{LV} grid states are highly underdetermined due to low real-time meter device coverage (low observability). 
The resulting uncertainty complicates forecasting of flexibility needs and requires \glspl{DSO} to request larger flexibility capacities, which increases costs. 

\subsubsection{Uncertainty about power system states due to frequent flexibility activations} \label{subsubsec:flexibility_uncertainty}
\Glspl{DSO} request and activate flexibility to avoid or postpone expensive grid extensions. 
However, frequent activations may break correlation between the few available measurements (e.g. primary substation and weather data) and system states at the end of \gls{LV} feeders \cite{muller2021uncertainty}. Thus, \gls{FM} operation might deteriorate the accuracy of \gls{LV} state estimation, making critical states potentially unobservable to \glspl{DSO}.
Based on inaccurate state estimations a \gls{DSO} might activate unnecessary or even counteracting flexibility, resulting in financial costs. 
In severe cases, the triggering of protection mechanisms might cause disconnection of customers. 

\subsubsection{Parallel events resulting in sudden change of flexibility needs} \label{subsubsec:parallel_events}
Different events, including line failures or shut down of large industrial loads, can lead to sudden change of the \glspl{DN} condition. 
Additionally, load peaks from simultaneous \gls{EV} charging and other new events will be introduced to \glspl{DN} in the upcoming years. 
If they occur during flexibility activation periods, such events may change grid condition in a way that activation is not required or even critical.
Moreover, state-sponsored actors could launch attacks on other systems, e.g. large battery energy storage systems or industrial plants, during activation periods to modify the grid condition.
Due to the low observability of \glspl{DN}, the detection of such events may be challenging. 

\subsection{Flexibility offer-based scenarios}
To sell flexibility, aggregators submit flexibility offers to \glspl{FM}. Depending on the market scheme, flexibility can be offered from intraday to months ahead. 

\subsubsection{Place wrong flexibility offers on the FM} \label{subsubsec:wrong_flex_offers}
If offers on the market do not reflect the actual potential, flexibility activations will likely not match the problem to solve. 
State-sponsored actors or insiders could tamper offers or place wrong offers on the market in the name of verified market participants. 
In less serious cases aggregators will have to pay a refund.
In severe cases critical grid conditions might not be solved by wrong flexibility offers.

\subsubsection{High uncertainty in the determination of flexibility offers} \label{subsubsec:Flex_offer_uncertainty}
Determination of flexibility potential is subject to uncertainties. The capacity of an aggregator portfolio is dependent on the comfort requirements of customers, weather, customer behavior and other portfolio changes. 
In particular, weather and customer behavior uncertainties directly translate into uncertainty of flexibility offers.
Moreover, in most cases the demand of small flexibility assets is controlled indirectly, e.g by adjusting temperature setpoints.
As the translation of temperature setpoints to power consumption is dependent on external factors, additional uncertainties are introduced during activation. Unreliable offers mainly reduce financial profit for aggregators. 
However, severe uncertainties might make the use of flexibility for \glspl{DSO} unreliable, and lead to more expensive but reliable alternatives, such as grid extensions. 
In case a \gls{DSO} relies on a flexibility offer to solve a critical condition, high uncertainty might result in disconnection of end users. 

\subsection{Flexibility measurement or schedule-based scenarios}
Reliable measurements of flexibility assets are required for service planning, activation and verification. Besides \gls{SM} readings, additional data may come from devices such as photovoltaic meters. 
To define the activation process, aggregators and flexibility asset owners agree on flexibility schedules. 

\subsubsection{Disrupt or manipulate flexibility measurements and schedules} \label{subsubsec:modify_flex_measurements_and_schedules}
Several actors might have an interest in manipulating flexibility measurements and schedules either by gaming or data tampering. 
Aggregators or flexibility asset owners could manipulate flexibility activation recordings for financial gain. 
Exemplary, for baseline services an asset owner could increase consumption before an activation period, to imitate a service by just returning to normal consumption level. 
Cyber-criminals that can sniff and modify data in networks of aggregators could compromise measurements, e.g. for blackmailing. 
One way is the modification of flexibility portfolio recordings to disrupt the service verification process. As a result, aggregators might receive fines for not fulfilling contractual agreements. 
Attackers could also modify the schedules which aggregators send to the assets, resulting in wrong activations. 
In mild cases, aggregators will be fined. In severe cases, wrong activations might trigger grid protection, resulting in disconnection of customers and thus high social costs. 

\subsection{Flexibility asset-based scenarios}
Flexibility assets comprise a variety of \glspl{DER}, owned by end users or companies. They reach from small loads such as refrigerators to large loads, including industrial processes. 

\subsubsection{Unavailability of flexibility assets} \label{subsubsec:flex_asset_unavailability}
During activation periods assets may not be available due to software failures, manual setpoint altering by asset owners or unforeseeable changes in the physical process of industrial flexibility assets. 
Moreover, cyber-criminals or state-sponsored actors could disturb communication by denial-of-service attacks. 
Since asset owners break the contract in cases of a failed activation, such scenarios would result in a financial penalty.
Especially in case of large flexibility assets, unavailability might lead to unresolved congestions and voltage violations.

\subsection{Vendor soft- and hardware-based scenarios} \label{subsec:vendor}
All actors of \glspl{FM} are dependent on services of third-parties, such as vendors. 
The required trust introduces potential risks.  

\subsubsection{Compromise vendor software and systems} \label{subsubsec:compromised_vendor}
State-sponsored actors could install malicious code in vendor soft- or hardware.
Attackers may install a backdoor in a \gls{PLC}.
This backdoor can later be used to manipulate \gls{DSO} operation in many ways.
The impact of such events may go beyond single end users, as \gls{EV} or \gls{HP} vendors provide soft- or hardware to multiple asset owners. 
The recent SolarWinds hack demonstrates the severity of such attacks \cite{peisert2021perspectives}. 
\section{Unresolved monitoring requirements for secure DSO participation in FMs} \label{sec:security_requirements}
\begin{figure}[t]
\centering
\includegraphics[width=1\linewidth,trim=18 18 190 19,clip]{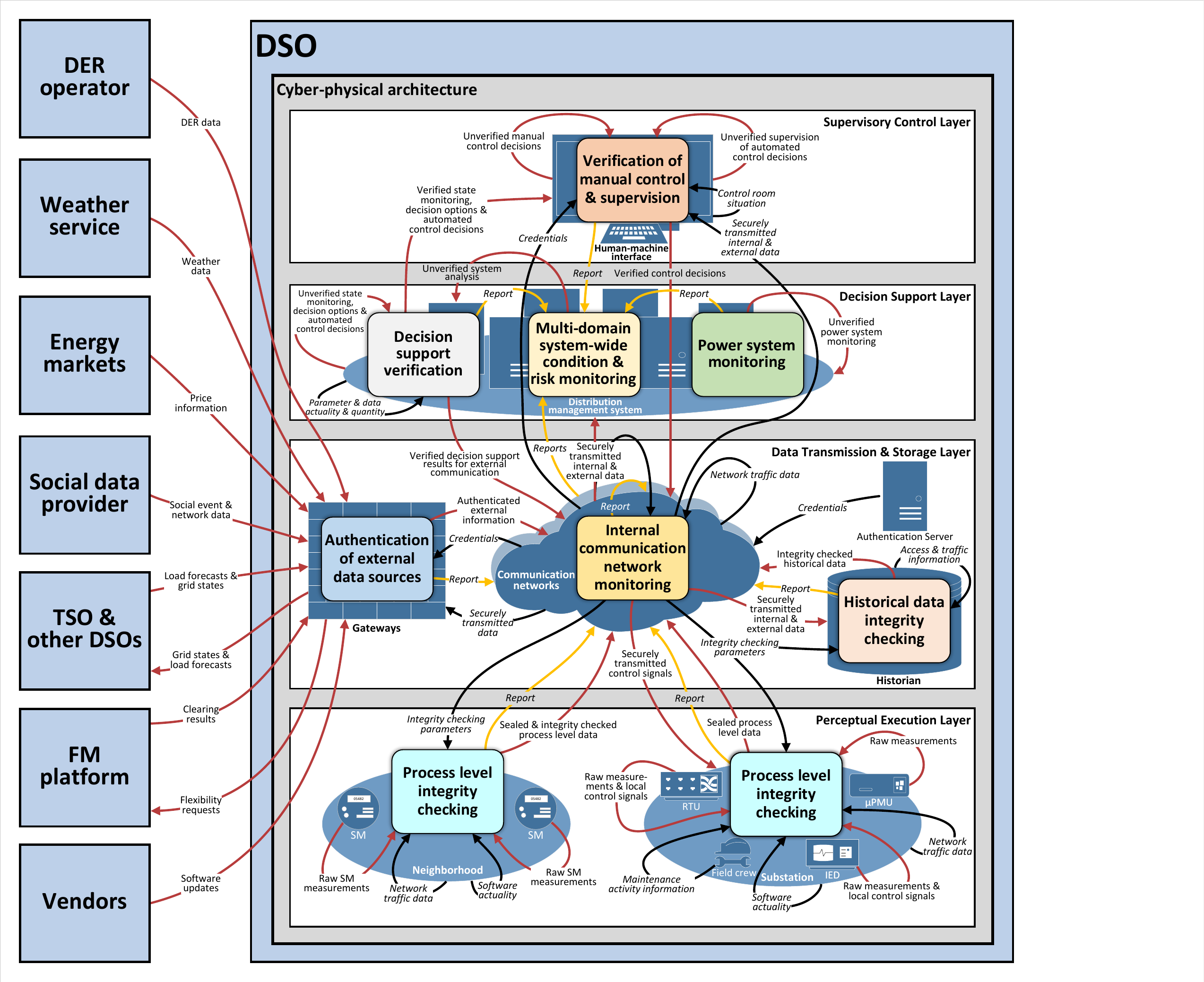}
\caption{Cyber-physical monitoring architecture of a \gls{DSO}. Colored fields: generic monitoring requirements, which are considered a prerequisite to observe the \gls{DSO}'s \gls{CPS}. Red links: information for power system operation. Black links: information for fulfilling monitoring requirements. Yellow links: reports of monitoring solutions satisfying the requirements.}
\label{fig:Framework}
\end{figure}

This section identifies unresolved monitoring requirements for \glspl{DSO} participating in \glspl{FM}
For that purpose, threat scenarios from Section \ref{sec:threat_scenarios} are mapped onto a generic cyber-physical monitoring architecture of \glspl{DSO}, shown in Fig. \ref{fig:Framework}.

\subsubsection{Quantifying flexibility-induced uncertainty}
Threat scenario \ref{subsubsec:flexibility_uncertainty} discusses that frequent flexibility activations could introduce uncertainty to \gls{LV} state estimation. 
At the same time, flexibility is used to operate power systems closer to capacity limits. 
Under these conditions, deterministic point estimations may fail silently, potentially impacting critical decisions.
On the contrary, probabilistic approaches provide information about reliability of estimates.
Incorporating such uncertainty quantification into the decision making process allows situational  adjustment of control actions and thus to lower the risk for wrong actions while retaining efficiency. 
Thus, quantifying flexibility-induced uncertainty is seen as important requirement for power system monitoring (Fig. \ref{fig:Framework}).

Uncertainty quantification also improves verification of data-driven decision support tools (Fig. \ref{fig:Framework}). 
Models that increase uncertainty under appearance of unseen flexibility activation events or system states provide operators with requirement indicators for retraining or additional input features \cite{muller2021uncertainty}.


Finally, uncertainty quantification facilitates analyzing the system-wide \gls{CPS} condition (Fig. \ref{fig:Framework}). If critical power system states are reported to a system-wide multi-domain condition monitor, probability of occurrence could be included. The improved interpretability could reduce false alarms, enabling more reliable system-wide monitoring of a \gls{DSO}'s \gls{CPS}.

\subsubsection{Flexibility activation detection}
Several threat scenarios (\ref{subsubsec:flexibility_asset_controller_modification}, \ref{subsubsec:unintentional_wrong_activation}, \ref{subsubsec:parallel_activations}) demonstrate that flexibility can be activated without the \gls{DSO} being aware of it. 
Such activations might occur intentionally through other market participants and cyber attackers or unintentionally due to human errors.
To enable immediate counteractions in case of critical activations, early detection is required.
Moreover, early detection would allow online verification of successful activation of DSO-requested flexibility by the operator \cite{muller2021unsupervised}.
Thus, for power system monitoring (Fig. \ref{fig:Framework}) automated real-time detection of flexibility activations is seen as an important requirement. 

\subsubsection{Flexibility scenario monitoring}
In threat scenario \ref{subsubsec:Flex_demand_uncertainty} and \ref{subsubsec:flexibility_uncertainty}, respectively, the difficulty of determining flexibility needs and flexibility activation demand is described. 
Especially under the aforementioned uncertainty and low observability of \glspl{DN}, flexibility planning becomes a challenging task for \glspl{DSO}. 
Thus, tools providing probabilistic power system state scenarios under various flexibility services are considered an important requirement for power system monitoring (Fig. \ref{fig:Framework}).
Depending on the market concept, tool requirements may look different.
For day or week-ahead procurement, tools will be required to provide state forecasts under various flexibility services. 
Market concepts that include real-time procurement of flexibility require tools for mapping available flexibility offers onto the current grid state.

\subsubsection{Integration of multi-domain information}
Many threat scenarios demonstrate a strong \gls{FM}-induced interaction and dependency among cyber, physical and market domains.
Cyber attackers may intend to cause physical damage (\ref{subsubsec:flexibility_asset_controller_modification}, \ref{subsubsec:modification_of_activation_signals}, \ref{subsubsec:modified_historical_data}) or disturb market actions (\ref{subsubsec:SM_modification}, \ref{subsubsec:wrong_flex_offers}, \ref{subsubsec:modify_flex_measurements_and_schedules}),
while insufficient coordination (\ref{subsubsec:parallel_activations}) or wrong flexibility offers (\ref{subsubsec:wrong_flex_offers}) on the \gls{FM} platform may result in physical impact. 
This interdependency has two consequences: on the one hand, underlying events are likely to leave traces in multiple domains. 
On the other hand, the root cause of a specific event can lay in different domains.
As an example, a denial-of-service attack against activation of a large flexibility asset
leaves traces in physical measurements and cyber network data.
Moreover, the activation failure could also be caused by a hardware failure or human error.
Thus, a monitoring requirement is seen in the integration of information from multiple domains to i) incorporate all available traces and ii) take possible threat origins in various domains into account. 
Among others, this somewhat general requirement could facilitate process-level or historical data integrity checking (Fig. \ref{fig:Framework}). 
One example is the integrated detection and classification of cyber attacks and physical faults by fusion of cyber network and physical process data \cite{muller2022assessment}. 
A central challenge for integration of multi-domain information is seen in the fusion of heterogeneous data. 

\subsubsection{Interpretable unsupervised intrusion and anomaly detection for flexibility assets}
In threat scenario \ref{subsubsec:multiple_SM_modification} and \ref{subsubsec:flexibility_asset_controller_modification}, respectively, it is demonstrated that edge devices, such as \glspl{SM} and \glspl{HEMS}, have security weaknesses (e.g., static encryption keys) which can be exploited by cyber attackers. 
\Glspl{FM} will make power system operation partly dependent on such less protected devices.  
Thus, from the perspective of the \gls{DSO}, advanced intrusion and anomaly detection systems for flexibility assets are considered as an important requirement for process level integrity checking (Fig. \ref{fig:Framework}). 
Challenges include computational constraints, lack of data describing the various attacks and anomalies, and the multitude of anomalies (e.g. cyber attacks, soft- and hardware faults and human errors) complicating root cause analysis.
A potential approach is seen in machine learning-based unsupervised anomaly detection on information stream level.
Unsupervised models do not require observations of anomalies.
Moreover, detecting anomalies on information stream level (e.g. destination IP addresses, customer setpoints and power demand) instead of system-wide, retains interpretability for root cause analysis also in an unsupervised scheme.
\section{Conclusion} \label{sec:conclusion}
In this work, threat scenarios for the \glspl{CPS} of \glspl{FM} are systematically formulated and presented. 
17 scenarios across all system domains are introduced, revealing several remaining security challenges. 
Among others, scenarios include simultaneous control of multiple flexibility assets by cyber attackers exploiting weak encryption, and uncertainty in the determination of flexibility needs and offers due to low meter coverage and high load variability in \glspl{DN}. 
Based on the threat scenarios, unresolved monitoring requirements for secure participation of \glspl{DSO} in \glspl{FM} are identified. 
Requirements include interpretable unsupervised anomaly detection for flexibility assets on information stream level and quantification of flexibility-induced uncertainty. 
By identifying such unresolved monitoring requirements, a foundation for new technical concepts and case studies addressing these gaps is provided.


\bibliographystyle{ieeetr}
\bibliography{07_References}

\begin{thebibliography}{10}

\bibitem{spiliotis2016demand}
K.~Spiliotis, A.~I. {Ramos Gutierrez}, and R.~Belmans, ``Demand flexibility
  versus physical network expansions in distribution grids,'' {\em Applied
  Energy}, vol.~182, pp.~613--624, 2016.

\bibitem{jin2020local}
X.~Jin and Q.~Wu, ``Local flexibility markets: Literature review on concepts,
  models and clearing methods,'' {\em Applied Energy}, vol.~261, 2020.

\bibitem{olivella2018optimization}
{P. Olivella-Rosell et al.}, ``Optimization problem for meeting distribution
  system operator requests in local flexibility markets with distributed energy
  resources,'' {\em Applied energy}, vol.~210, pp.~881--895, 2018.

\bibitem{alizadeh2016flexibility}
{M. Alizadeh et al.}, ``Flexibility in future power systems with high renewable
  penetration: A review,'' {\em Renewable and Sustainable Energy Reviews},
  vol.~57, pp.~1186--1193, 2016.

\bibitem{sperstad2020impact}
I.~B. Sperstad, M.~Z. Degefa, and G.~Kjølle, ``The impact of flexible
  resources in distribution systems on the security of electricity supply: A
  literature review,'' {\em Electric Power Systems Research}, vol.~188, 2020.

\bibitem{zeng2017impact}
B.~Zeng, G.~Wu, J.~Wang, J.~Zhang, and M.~Zeng, ``Impact of behavior-driven
  demand response on supply adequacy in smart distribution systems,'' {\em
  Applied Energy}, vol.~202, pp.~125--137, 2017.

\bibitem{ghose2019risk}
T.~Ghose, H.~W. Pandey, and K.~R. Gadham, ``Risk assessment of microgrid
  aggregators considering demand response and uncertain renewable energy
  sources,'' {\em Journal of Modern Power Systems and Clean Energy}, vol.~7,
  no.~6, pp.~1619--1631, 2019.

\bibitem{chehri2021security}
A.~Chehri, I.~Fofana, and X.~Yang, ``Security risk modeling in smart grid
  critical infrastructures in the era of big data and artificial
  intelligence,'' {\em Sustainability}, vol.~13, no.~6, 2021.

\bibitem{smart-meter}
M.~Costache and V.~Tudor, ``Security aspects in the advanced metering
  infrastructure,'' {M.Sc. Thesis}, Chalmers University of Technology,
  Department of Civil and Environment, Gothenburg, Sweden, 2011.

\bibitem{enisa}
L.~Marinos, ``Smart grid threat landscape and good practice guide,'' {\em White
  Paper, European Network and Information Security Agency (ENISA); ENISA:
  Attiki, Greece}, 2013.

\bibitem{8220480}
N.~Kshetri and J.~Voas, ``Hacking power grids: A current problem,'' {\em
  Computer}, vol.~50, no.~12, pp.~91--95, 2017.

\bibitem{alanazi2021load}
F.~Alanazi, J.~Kim, and E.~Cotilla-Sanchez, ``Load oscillating attacks of smart
  grids: Demand strategies and vulnerability analysis,'' {\em arXiv preprint
  arXiv:2105.00350}, 2021.

\bibitem{liu2011false}
Y.~Liu, P.~Ning, and M.~K. Reiter, ``False data injection attacks against state
  estimation in electric power grids,'' in {\em Proceedings of the 16th ACM
  Conference on Computer and Communications Security}, p.~21–32, 2009.

\bibitem{muller2021uncertainty}
N.~Müller, S.~Chevalier, C.~Heinrich, K.~Heussen, and C.~Ziras, ``Uncertainty
  quantification in {LV} state estimation under high shares of flexible
  resources,'' {\em Electric Power Systems Research}, vol.~212, 2022.

\bibitem{peisert2021perspectives}
{S. Peisert et al.}, ``Perspectives on the solarwinds incident,'' {\em IEEE
  Security Privacy}, vol.~19, no.~2, pp.~7--13, 2021.

\bibitem{muller2021unsupervised}
N.~Müller, C.~Heinrich, K.~Heussen, and H.~W. Bindner, ``Unsupervised
  detection and open-set classification of fast-ramped flexibility activation
  events,'' {\em Applied Energy}, vol.~312, 2022.

\bibitem{muller2022assessment}
N.~Müller, C.~Ziras, and K.~Heussen, ``Assessment of cyber-physical intrusion
  detection and classification for industrial control systems,'' in {\em 2022
  IEEE International Conference on Communications, Control, and Computing
  Technologies for Smart Grids (SmartGridComm)}, 2022.

\end{thebibliography}
\end{document}